\documentclass[manuscript]{acmart}
\AtBeginDocument{%
  }
\usepackage{array} 
\usepackage{graphicx} 
\usepackage{float} 
\usepackage{ragged2e} 
\usepackage{booktabs} 
\usepackage{multirow} 
\usepackage{makecell} 
\setcopyright{cc}

\copyrightyear{2025}
\acmYear{2025}
\title{Promoting Real-Time Reflection in Synchronous Communication with Generative AI}
\author{Yi Wen}
\affiliation{%
  \institution{Texas A\&M University}
  \country{USA}
}
\email{cyberwenyi2357@tamu.edu}

\author{Meng Xia}
\affiliation{%
  \institution{Texas A\&M University}
  \country{USA}
}
\email{mengxia@tamu.edu}
\renewcommand{\shortauthors}{Wen et al.}
\begin{document}
\begin{abstract}
 Real-time reflection plays a vital role in synchronous communication. It enables users to adjust their communication strategies dynamically, thereby improving the effectiveness of their communication. Generative AI holds significant potential to enhance real-time reflection due to its ability to comprehensively understand the current context and generate personalized and nuanced content. However, it is challenging to design the way of interaction and information presentation to support the real-time workflow rather than disrupt it. In this position paper, we present a review of existing research on systems designed for reflection in different synchronous communication scenarios. Based on that, we discuss design implications on how to design human-AI interaction to support reflection in real time.
\end{abstract}
\maketitle
\begingroup
\renewcommand\thefootnote{}\footnote{
 This paper was presented at the 2025 ACM Workshop on Human-AI Interaction for Augmented Reasoning (AIREASONING-2025-01). This is the authors’ version for arXiv.}
\endgroup
\section{Introduction}
The definition of reflection in HCI field, as suggested by Eric et al.\cite{baumer2014reviewing}, is ``reviewing a series of previous experiences, events, etc., and putting them together in such a way as to come to a better understanding or gain some sort of insight''. It is considered a central process in many fields, such as healthcare\cite{luo2021foodscrap,cho2021reflect}, education\cite{zhang2023vizprog}, and design\cite{choi2024creativeconnect}. In HCI field, there is a significant body of research on the design for reflection\cite{baumer2014reviewing}. However, they often focus on fostering reflection after the fact rather than enabling reflection in real time. Real-time reflection, on the other hand, is particularly useful in synchronous communication scenarios\cite{aseniero2020meetcues,chen2025meetmap,ma2022glancee,sun2019presenters} as it allows individuals to consciously evaluate and adapt their communication rather than recalling the entire process afterwards \cite{samrose2021meetingcoach,samrose2018coco}. 

Despite its potential benefits, real-time reflection is inherently challenging for individuals. In synchronous communication, speakers need to actively contribute to the ongoing conversation, leaving little cognitive bandwidth to reflect on past interactions. Moreover, in many cases, it is difficult for the speaker to realize when their communication is unclear or misaligned, since there is a lack of immediate feedback from the audience. For example in contexts like remote communication, some nonverbal communication cues will be hindered, making it hard for the speaker to gauge the audience's reactions. In solo practice, it is also hard to reflect in real time, as there is no audience. 

Recently, there has been a growing body of work that uses technological intervention to support reflection during the activity. In educational settings, real-time reflection has been specifically targeted to improve instructor-learner interactions and enhance teaching strategies. For example, Glancee \cite{ma2022glancee} developed a comprehensive learning status detection algorithm to help instructors grasp the learning status of students in online synchronous classes. EduLive \cite{edulive} achieved a similar purpose by aggregating learners' transcript-based annotations. There is another line of work focus on promoting reflection in online meeting. For example, TalkTraces \cite{chandrasegaran2019talktraces} captures and visualizes verbal content in meetings to help participants reflect on conversational dynamics. MeetMap\cite{chen2025meetmap} uses large language models (LLMs) to provide mapping of collaborative dialogue in real-time in online meetings to help participants track the structure and flow of conversations, which is the first step in reflection\cite{li2011understanding}. Moreover, Joshua \cite{mcveigh2018immersive} brings the speech visualization mechanic into VR space to help participants reflect on conversational imbalances. Beyond these, real-time reflection in practice settings has also been explored to help individuals refine their skills. TutorUp\cite{pan2025tutorup} utilizes LLM to provide immediate and personalized feedback to help novice tutors better engage with their students. AudiLens\cite{10.1145/3586182.3625114} leverages LLMs to simulate diverse personas of the audience and generate real-time feedback. 

In this position paper, we review existing systems designed for supporting reflection in synchronous communication. Some only support post-activity reflection while most of them support during-the activity reflection. We conclude the design patterns of existing systems and discuss how Generative AI can further shift this pattern. Based on that, we discuss the design implications for using Generative AI to scaffold real-time reflection in synchronous communication.

\section{Method}
Given our precise focus on supporting reflection in synchronous communication to enhance the communication strategy and performance, we first identified common synchronous communication scenarios where reflection is especially useful. These scenarios include remote meeting, lectures, presentations, as well as practice talks.
We searched in the ACM Digital Library using keywords 'meeting','reflection','online classes', 'presentation', 'practice', and 'training'. To ensure that our review captures the most recent trends in this area, we filtered the collected papers based on their year of publication, focusing only on those published in the last five years.

Finally, there are 11 papers that satisfy the conditions. We conclude on ways to support reflection, the role of Generative AI in the system, and the interaction paradigms, which is mapped to design pattern and notification level. We referred to the taxonomy of ambient information systems proposed by Zachary et al.\cite{pousman2006taxonomy} when analyzing the interaction paradigm.

\section{Results}
\newcolumntype{P}[1]{>{\raggedright\arraybackslash}p{#1}}
\begin{table}[htbp]
    \small
    \setlength{\tabcolsep}{4pt}
    \renewcommand{\arraystretch}{1.3}  
    \centering
    \caption{Overview of Real-Time Reflection Systems}
    \label{tab:realtime_reflection}
    \begin{tabular}{|p{3cm}|p{3.3cm}|p{3.3cm}|p{2.2cm}|p{3.5cm}|} 
    \hline
    \multicolumn{2}{|p{6.4cm}|}{\textbf{Ways to support real-time reflection}} & 
    \textbf{Design Pattern} &
    \textbf{Notification Level} &
    \textbf{The role of Generative AI} \\
    \hline

    Increasing user's contextual awareness & 
    Augmenting/Simulating Audience Feedback \cite{10.1145/3586182.3625114,aseniero2020meetcues} & 
    Information Monitor Display \cite{aseniero2020meetcues}, Multiple-Information Consolidators \cite{10.1145/3586182.3625114} & 
    Make Aware & 
    Understand the conversation and generate feedback as an audience \\

    \cline{2-5}
    & Aggregating Audience Status \cite{edulive,ma2022glancee,sun2019presenters,mcveigh2018immersive,samrose2018coco} & 
    Information Monitor Display \cite{edulive,ma2022glancee,samrose2018coco,sun2019presenters}, Multiple-Information Consolidators \cite{edulive}, Symbolic Sculptural Display \cite{mcveigh2018immersive} & 
    Make Aware \cite{edulive,ma2022glancee,samrose2018coco,mcveigh2018immersive}, Interrupt \cite{sun2019presenters} & 
    Traditional NLP/CV techniques are used, no need for Generative AI \\

    \cline{2-5}
    & Real-time Summary of Past Conversation \cite{chen2025meetmap,chandrasegaran2019talktraces} & 
    Multiple-Information Consolidators \cite{chen2025meetmap,chandrasegaran2019talktraces}, High Throughput Textual Display \cite{chen2025meetmap} & 
    Change Blind \cite{chandrasegaran2019talktraces}, Demand Attention \cite{chen2025meetmap} & 
    Identify topics, summarize, and analyze the conversation \\

    \hline
    Evaluating user's performance and providing actionable suggestions & 
    Providing Expert Suggestion \cite{pan2025tutorup} & 
    High Throughput Textual Display & 
    Demand Attention & 
    Understand the conversation and generate suggestion as an expert \\
    \hline
    \end{tabular}
\end{table}

\subsection{How (Real-Time) Reflection of Synchronous Communication is Supported?}
We concluded the ways to support reflection in synchronous communication. We divide them into two categories, one is supporting reflection through \textbf{increasing user's contextual awareness}, and the other is \textbf{evaluating user's performance and providing actionable suggestions}, as shown in the first column of Table ~\ref{tab:realtime_reflection} .

The first three rows in Table~\ref{tab:realtime_reflection} belong to the first category. In synchronous communication scenarios, it can be challenging for the speaker to comprehensively understand the audience's status and feedback in real time. Therefore, using AI to analyze the current situation comprehensively and present the results to the speaker in a glanceable way could help them reflect on their performance. Additionally, providing a summary of past conversations can assist in reflection, as it allows the speaker to compare the discussion with their planned agenda, helping them stay on track and avoid deviating from the topic.

In addition to the methods mentioned above, providing professional feedback or suggestions from experts can also enhance reflection in synchronous communication, especially in domains that require specialized knowledge, such as teaching. In such scenarios, expert insights can offer valuable guidance, helping participants refine their approach and improve the quality of communication in real time.
\subsection{Interaction Paradigm}
The interaction paradigm for supporting reflection in synchronous communication largely depends on how and when information is delivered to users. Based on our review, we identify three key paradigms: user-initiated, system-initiated (proactive), and continuous display.

\textbf{User-initiated interaction} allows users to decide when to access reflective information, demanding user attention. For example, in systems like EduLive \cite{edulive}, users can manually check aggregated audience status or other analytics as needed. This paradigm gives users full agency but may result in extra cognitive efforts and missed opportunities for timely reflection.

\textbf{System-initiated interaction} leverages proactive notifications to deliver feedback or suggestions at critical moments without requiring user prompts, which make users aware of the information change with light notification, or interrupt users. For instance, TutorUp\cite{pan2025tutorup} and AudiLens \cite{10.1145/3586182.3625114} use Generative AI to actively monitor the communication context and provide timely suggestions or simulate audience feedback. This approach can reduce user workload while making them aware of the information being presented, but may risk interrupting the user's current workflow if not well designed.

\textbf{Continuous display} keeps real-time information persistently visible to users, enabling them to monitor and reflect on ongoing activities at a glance. These systems can be change blind. For example, TalkTraces\cite{chandrasegaran2019talktraces} continuously present conversation summaries with users. They can ignore the subtle changes in the visualizations when focusing on the conversation. In contrast, MeetScript\cite{chen2023meetscript} provided a parallel communication channel through high throughput textual display, which demands user attention to interact with the script. This approach supports ongoing awareness and reflection, though it may reduce user agency and potentially contribute to information overload if not properly managed.


\subsection{What Kind of Opportunities Generative AI Brings in Supporting Real-time Reflection?}
    From Table ~\ref{tab:realtime_reflection}, we can see that traditional AI is used mostly in aggregation of audience status relies on simple statistics (e.g., facial expression counts, hand-raising, voting results), often shown as basic pie or bar charts. These only reflect surface-level distributions. 

    Generative AI, like Large Language Models or Vison Language Models, could greatly enhance the aggregation of audience status and meeting content visualization by moving beyond basic statistics to generate nuanced, natural language summaries, infer conversation structures, extract key opinions and consensus points while integrating multimodal data (such as emotions, behaviors) for richer, cross-modal visualizations. 

Moreover, the generative capabilities and few-shot learning ability of Generative AI present new opportunities for real-time reflection. Specifically, LLMs can role-play as any participant in the communication context—for example, acting as a typical audience member\cite{10.1145/3586182.3625114}, or, when provided with domain knowledge, as an expert\cite{pan2025tutorup}. This flexibility allows Generative AI to deliver real-time, nuanced, and contextual feedback or suggestions from multiple perspectives, enhancing the reflection process in synchronous communication.
\section{Design Implications for Promoting Real-Time Reflection in Synchronous Communication with Generative AI}
 Based on our analysis of existing systems, we can see the significant potential of Generative AI to improve real-time reflection in synchronous communication scenarios. However, the current design is not perfect. We analyze the limitations of current systems based on the findings of the user study and propose the following design implications.

\subsection{Addressing the Lack of Interpretability}
One of the key challenges that users face when using systems to support real-time reflection is the lack of interpretability in AI-generated results. When system outputs deviate from user expectations, it can lead to confusion, distrust, and increased cognitive load. To address this, designers should consider incorporating lightweight explanations or visual indicators that clarify how AI generated specific results, without overwhelming users with excessive information. For example, including brief annotations or contextual hints on dashboards can help users build trust and better understand the system’s outputs.
\subsection{Leveraging Proactiveness to Reduce User Workload}
Users in synchronous communication scenarios often appreciate systems that proactively assist them, as long as the timing and relevance of proactive actions are appropriate. Generative AI can identify critical moments where users might need support and provide timely notifications or suggestions. This reduces the number of manual steps required for interaction and reduces the user workload. For example, proactive summarization of conversations or agenda tracking can help users focus on their primary tasks without having to search for information manually.
\subsection{Addressing the Limitations of Role-Play Agents in Synchronous Communication}
While role-playing agents powered by generative AI can enhance real-time feedback in synchronous communication, their effectiveness is limited by the quality of persona design and contextual grounding \cite{chen2025towards}. If personas are too generic or lack relevant context, the simulated behavior may become unconvincing, which could bring negative effect on user's reflection. Additionally, the flexibility to role-play any character can lead to inconsistencies or superficial responses, especially in complex or specialized scenarios. Designers should therefore emphasize richer, context-sensitive persona modeling and implement clear evaluation criteria to ensure that role-playing agents deliver believable and trustworthy support.


\bibliographystyle{ACM-Reference-Format}
\bibliography{sample-base}

\appendix

\end{document}